\documentclass[10pt,conference]{IEEEtran}
\usepackage{amsmath,amssymb,amsfonts}
\usepackage{graphicx}
\usepackage{textcomp}
\usepackage{xcolor}
\usepackage{cite}
\usepackage[hyphens]{url}
\usepackage[caption=false,font=footnotesize]{subfig}
\usepackage{booktabs}
\newif\ifshowauthornotes
\showauthornotestrue
\ifdefined\hideauthornotes
  \showauthornotesfalse
\fi
\definecolor{authornotecolor}{RGB}{150,35,120}

\begin{document}
\title{Choosing Coding-Agent Benchmarks for Practice:\\Lessons from Task Demands and Agent Behaviour}
\author{\IEEEauthorblockN{Radin Shayanfar}
\IEEEauthorblockA{\textit{Dept. of Electrical and Computer Engineering} \\
\textit{Queen's University}\\
Kingston, Canada \\
radin.shayanfar@queensu.ca}
\and
\IEEEauthorblockN{Keheliya Gallaba}
\IEEEauthorblockA{\textit{Centre for Software Excellence}\\
\textit{Huawei Canada}\\
Kingston, Canada \\
gallabak@sigsoft.org}
\and
\IEEEauthorblockN{Ahmed E. Hassan}
\IEEEauthorblockA{\textit{School of Computing} \\
\textit{Queen's University}\\
Kingston, Canada \\
ahmed@cs.queensu.ca}
}

\maketitle
\pagestyle{plain}
\thispagestyle{plain}
\begin{abstract}
Software teams evaluating coding agents need evidence that reflects the changes they expect those agents to make. Benchmark scores summarize how often an agent resolves tasks under broad labels such as bug fixing and feature implementation, but provide little guidance on whether the evaluated work resembles a team's workload. To support this decision, we characterize 2{,}487 tasks from five public benchmarks and analyze 14{,}922 runs of six agent configurations. We introduce the Spread--Novelty--Centrality (SNC) profile to describe how the reference solution is distributed across files and directories, how much it adds relative to what it removes, and the dependencies, history, and complexity of the code it touches. We also compare the scope of agent patches and their file exploration. Every pair of benchmarks differs on at least two SNC axes, including benchmarks with the same task label. Among tasks resolved by Claude Opus 4.6, patches are larger than the reference on FeatBench and smaller on FeatureBench. Across the six configurations, resolved runs tend to involve less distributed changes and code with lower values on the Centrality indicators, while the patch sizes associated with resolution differ between the Claude and Qwen configurations. These observations suggest that practitioners should select evaluation tasks based on their change characteristics, inspect the information provided in task descriptions, and assess patch scope alongside test outcomes. We explain how teams can apply this analysis to completed changes and identify the additional evidence needed before using it for task assignment or review automation.
\end{abstract}
\begin{IEEEkeywords}
coding agents, benchmark selection, software engineering practice, software maintenance, empirical software engineering
\end{IEEEkeywords}
\section{Introduction}
\begin{figure}[!t]
  \centering
  \includegraphics[width=\linewidth]{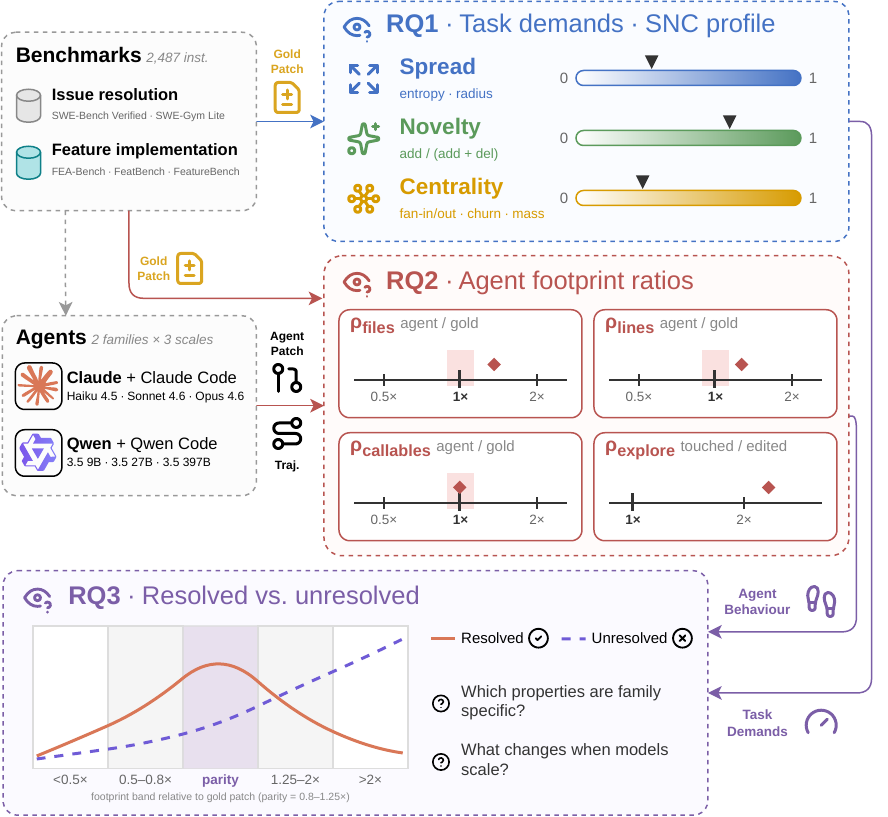}
  \caption{Overview of the study: the SNC profile characterizes task demands, while agent patches and trajectories support analyses of resolution and agent behaviour.}
  \label{fig:method}
\end{figure}
Coding agents combine a large language model (LLM) with tools for reading files, editing code, and running commands. Developers can give an agent a change request and ask it to produce a patch within an existing repository. Studies of agent contributions to open source projects show that these patches are entering development workflows, where maintainers still need to assess and revise them~\cite{use-agentic-coding-2026-watanabe,aidev-2025-hao-li}. Industrial deployments also span diverse workflows. Organizations have reported thousands of accepted changes across privacy automation, framework adoption, and development tasks, with acceptance varying across automation domains~\cite{whatscode-2026-mao}. For teams evaluating this technology, a practical question follows: which evaluation results provide useful evidence about the work they intend to delegate?

Public benchmarks evaluate coding agents on repository tasks and report the proportion that pass the prescribed tests~\cite{jimenez2024swebench,feabench-2025-li,featbench-2025-chen}. A team also needs to understand the work represented by that score. A small correction in one file and a change to a shared component may both be labelled a bug fix. Similarly, implementing a feature may involve adding a component or modifying behaviour across existing components. These distinctions matter when deciding what evidence to collect before adoption.

From the perspective of engineers selecting and interpreting evaluations, benchmark labels describe a change's intended purpose but reveal little about its distribution or relationship to surrounding code. Recent evidence also reinforces that gains on one coding task may not transfer to related tasks~\cite{jetbrains-shibaev2026dont}. In domain-specific code generation, strong general-domain performance likewise does not consistently imply strong performance in each application domain~\cite{domaincodebench-2026-zheng}. Benchmark construction adds another consideration. Some tasks provide the names and signatures of functions to implement; others require the agent to identify these itself. A team that evaluates only the former receives limited evidence about an agent's ability to work from requests that omit these details. Experiments with underspecified variants of SWE-bench Verified tasks further show that the information supplied to an agent, and its opportunity to obtain clarification, affect measured resolution~\cite{ambig-swe-2026-vijayvargiya}.

Studies of SWE-bench submissions show that patches accepted by benchmark tests can still fail broader developer tests~\cite{solved-issues-2026-wang}. Industry evaluation guidance also recommends examining agent transcripts and code quality alongside test outcomes~\cite{agent-evals-2026-grace}.

Our study began with the need to select a benchmark to evaluate the feature implementation capabilities of coding agents. Publicly available benchmarks differed in the pull requests (PRs) they were built from, and the implementation hints they supplied, leaving us uncertain about which capabilities each measured and which to choose. These qualitative comparisons motivated us to quantify task characteristics.

To make these differences measurable, we introduce the \emph{Spread--Novelty--Centrality (SNC) profile}. Spread describes how a change is distributed across files and directories. Novelty describes the balance between added and removed code. Centrality describes the dependencies, recent modification history, and complexity of the affected code. We compute these characteristics from the benchmark's reference solution, commonly called the \emph{gold patch}. We then examine the agent's own patch and its \emph{trajectory}, the record of its actions during a run. This second analysis captures differences that a reference solution alone cannot show.

Our study covers 2{,}487 tasks from five benchmarks and 14{,}922 agent runs. We examine two families, Claude and Qwen, with three configurations in each family. The study addresses three questions that support different parts of an evaluation decision. Figure~\ref{fig:method} summarizes how the SNC profile and behavioural measures support the three research questions.

\paragraph{RQ1: Do benchmarks with the same task label represent similar changes?}
\textit{Motivation:} Teams need to know whether adding another benchmark expands the kinds of work they evaluate. \textit{Results:} Every pair of benchmarks differs on at least two SNC axes, including pairs with the same broad task label. Several differences are consistent with the benchmarks' construction rules. \textit{Implication:} Select benchmarks by the changes they contain and compare their profiles with completed changes from the intended workload.

\paragraph{RQ2: How do an agent's patches and exploration differ across benchmarks?}
\textit{Motivation:} A passing test result does not describe the scope of the resulting change or the code the agent inspected. \textit{Results:} On the 1{,}083 tasks Claude Opus 4.6 resolves, its patches are larger than the gold patches on FeatBench and smaller on FeatureBench. We detect no separation between benchmarks in exploration breadth under our statistical test. \textit{Implication:} Inspect patch scope and the hints supplied by each benchmark when interpreting resolution rates.

\paragraph{RQ3: Which task and behaviour characteristics are associated with resolution?}
\textit{Motivation:} A useful evaluation should identify where agents struggle and whether those patterns hold across configurations. \textit{Results:} Resolved runs tend to involve lower Spread and Centrality across both families. The relationship with Novelty is not monotonic. Patch scope differs by configuration: larger Claude configurations more often match the gold patch's scope, while resolved Qwen runs frequently exceed it. \textit{Implication:} Examine task characteristics across configurations, and establish configuration-specific baselines before using patch size as a diagnostic.

We contribute a method for examining evaluation coverage, evidence from five benchmarks, and lessons for selecting tasks and inspecting agent changes. Uses such as assigning new requests to agents require further validation in development workflows.

\section{Background and Related Work}
\label{sec:background}
\subsection{What a Coding-Agent Benchmark Evaluates}
A repository task starts from a \emph{base commit}, the version of the code before the requested change. The agent receives this code and a natural-language problem statement. Its patch records the lines it adds and removes. The benchmark then executes its prescribed tests to determine whether the task is resolved. The gold patch provides a reference implementation, but other patches may pass the same tests. Throughout this paper, \emph{resolved} means that the benchmark accepts the run; it does not establish that maintainers would accept the patch or that all relevant behaviour is tested~\cite{openai2026swebenchverified,openai2026codingevaluations}.

A model is only one part of an agent configuration. The surrounding agent software decides how to present tools, retain context, and execute model requests. We use the term \emph{agent scaffold} to refer to this software, such as Claude Code~\cite{claude-code-anthropic-2025} or Qwen Code~\cite{qwen-code-2025}. We use \emph{evaluation harness} to refer to the infrastructure that prepares benchmark tasks, launches agents, and checks outcomes. Keeping this distinction explicit is necessary when attributing differences between configurations.

\subsection{The Five Studied Benchmarks}
\label{sec:categorization}
Repository benchmarks cover tasks ranging from issue resolution to feature implementation~\cite{agentic-swe-survey-2025-guo,agentic-issue-res-survey-2025-jiang}. We study two benchmarks for issue resolution and three for feature implementation. Their construction rules provide different forms of evidence about agent capabilities.

\textbf{SWE-bench} pairs GitHub issues with merged PRs and tests that fail before the change and pass afterward~\cite{jimenez2024swebench}. We use SWE-bench Verified, a subset screened by humans for task specification and test reliability~\cite{chowdhury2024swebenchverified}. \textbf{SWE-Gym} uses a similar task formulation, with repositories disjoint from SWE-bench, and was developed as a training environment~\cite{swegym-2024-pan}. We use its Lite subset as a second collection of issue-resolution tasks.

\textbf{FEA-Bench} selects PRs that add components such as functions or classes. It supplies signatures and docstrings as hints so that implementations can be evaluated through the expected interfaces~\cite{feabench-2025-li}. \textbf{FeatBench} selects PRs that modify existing functions without adding or deleting functions. An LLM rewrites PR information into a problem statement without implementation hints~\cite{featbench-2025-chen}. \textbf{FeatureBench} constructs tasks by removing source lines covered by selected tests. Its construction requires at least 100 removed lines and 10 tests that fail before restoration and pass afterward. Most tasks in its full subset retain interface signatures as stubs~\cite{featurebench-2026-zhou}. Thus, the three collections use the term feature implementation for different kinds of change.

\begin{figure}[t]
  \centering
  \includegraphics[width=\linewidth]{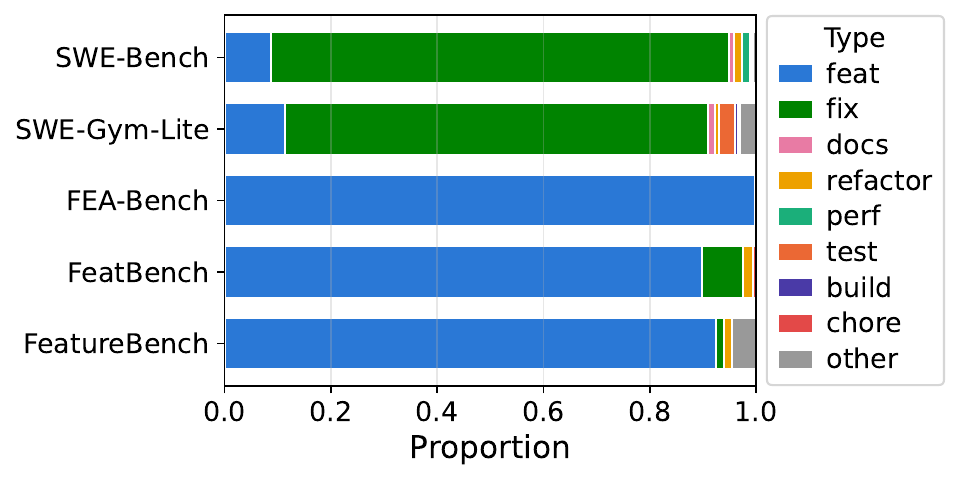}
  \caption{Problem-statement labels under the Conventional Commits taxonomy for the five studied benchmark subsets.}
  \label{fig:problem-classification}
\end{figure}

To compare task labels uniformly, we classify each problem statement under the Conventional Commits taxonomy~\cite{conventionalcommits}, following the LLM classification approach of Li et al.~\cite{aidev-2025-hao-li}. Figure~\ref{fig:problem-classification} shows that \texttt{fix} dominates SWE-bench Verified and SWE-Gym Lite, while \texttt{feat} dominates the three feature benchmarks. FeatBench and FeatureBench have similar label distributions despite their different construction rules. These automated labels describe task purpose and may contain classification errors; the SNC profile examines the reference change itself (Section~\ref{sec:rq1}).

\subsection{Interpreting Benchmark Evidence}
Studies of benchmark quality examine whether an evaluation measures the intended capability~\cite{bean2025measuring,reuel2024betterbench}. For coding agents, concerns include contamination from benchmark data~\cite{swebench-illusion-2025-liang}, differences between benchmark descriptions and developer requests~\cite{saving-swe-2025-garg}, weak test suites~\cite{swe-abs-2026-yu}, and variation between repeated runs~\cite{randomness-agentic-2026-bjarnason}. Our analysis addresses a complementary question: what kinds of code changes contribute to a reported score?

This question also matters when interpreting progress during model development. Shibaev et al.~\cite{jetbrains-shibaev2026dont} report limited transfer from optimization on one coding benchmark to other evaluations. Our profile provides a way to inspect whether evaluation collections differ in their change characteristics. It does not establish the cause of transfer failures. Similarly, studies of agent PRs examine what developers accept in practice~\cite{use-agentic-coding-2026-watanabe}. We examine benchmark acceptance and patch scope, which are inputs to such an assessment but do not replace evidence about maintainability or developer effort.

Chen et al.~\cite{beyond-final-code-2026-chen} analyze SWE-bench trajectories and testing logs, finding that execution errors correlate with lower resolution rates and increased reasoning overhead. Their process-oriented error analysis complements our examination of patch scope and file exploration across benchmark collections.

\section{Study Design}
\label{sec:study-design}
\subsection{Evaluation Context and Data}
\label{sec:study-benchmarks}
We use public tasks so that the comparison can be repeated without access to proprietary repositories. Table~\ref{tab:study-benchmarks} lists the 2{,}487 instances. These tasks provide a common basis for comparing configurations, but they are not a sample of the work performed by a particular company. FEA-Bench contributes 1{,}401 instances and therefore has more influence on pooled analyses than the other benchmarks. We report benchmark comparisons separately in RQ1 and RQ2 and interpret the pooled results of RQ3 with this imbalance in mind.

\begin{table}[t]
  \centering
  \caption{Studied tasks and information supplied to the agent. FeatureBench provides hints for most tasks in its full subset.}
  \label{tab:study-benchmarks}
  \begin{tabular}{@{}lrl@{}}
    \toprule
    Benchmark & Tasks & Description source; hints \\
    \midrule
    SWE-bench Verified & 500 & GitHub issue; no \\
    SWE-Gym Lite & 230 & GitHub issue; no \\
    FEA-Bench & 1{,}401 & PR metadata; yes \\
    FeatBench & 156 & PR metadata; no \\
    FeatureBench & 200 & Test coverage; yes \\
    \midrule
    Total & 2{,}487 & \\
    \bottomrule
  \end{tabular}
\end{table}

The unit of analysis in RQ1 is a task and its gold patch. In RQ2 and RQ3 it is an agent run on a task. A task's SNC profile is shared by its six runs because the reference change is fixed. Agent patches, trajectories, and outcomes vary by configuration. We therefore treat these as 2{,}487 tasks evaluated under six configurations.

\subsection{Agent Configurations and Execution}
\label{sec:study-agents}
We evaluate Claude Haiku 4.5, Sonnet 4.6, and Opus 4.6 through Claude Code~\cite{claude-code-anthropic-2025}. We evaluate Qwen 3.5 9B, 27B, and 397B-A17B through Qwen Code~\cite{qwen3.5}. We refer to these as small (S), medium (M), and large (L) within each family. The labels indicate the configuration ordering used in this study; they do not equate parameter counts or computational budgets across the two families. Claude Opus 4.6 is the highest-performing configuration in our study and is the reference configuration for RQ2.

\label{sec:study-harness}
All runs use Harbor~\cite{Harbor_Framework,shi2026harbor} as the evaluation harness. We converted FEA-Bench tasks into Harbor's format because Harbor did not natively support FEA-Bench. The experiment comprises 30 combinations of benchmark and agent configuration and 14{,}922 trajectories, corresponding to the six configurations across the 2{,}487 tasks. Harbor provides common orchestration, while Claude Code and Qwen Code remain distinct agent scaffolds.

Execution records span April 14--June 2, 2026. Harbor used Docker environments. We used the default timeouts specified in the Harbor task definitions. The scaffold versions used are Claude Code 2.1.108 and Qwen Code 0.14.4. We use the agents' default decoding parameters.

\subsection{Measurements and Analysis}
Gold patches and repository structure provide the SNC indicators. Agent patches and recorded file accesses provide measurements of scope and exploration. Benchmark outcomes allow us to contrast resolved and unresolved runs.

The SNC indicators describe completed changes and can support retrospective evaluation of historical work. Assigning new requests to agents would require estimating the affected code, which we do not evaluate. Comparisons with a gold patch are also generally unavailable for new requests.

We use Radon~\cite{radon} to identify code blocks overlapping patch line ranges. Imports resolve to the longest matching repository module prefix, excluding external, unresolved, and self-imports.

\section{RQ1: Do Benchmarks with the Same Task Label Represent Similar Changes?}
\label{sec:snc}
\label{sec:rq1}
\subsection{Motivation}
We apply the SNC profile to all 2{,}487 gold patches to examine whether benchmarks with the same task label cover similar changes. The profile groups seven observable indicators under three axes. It does not directly measure developer expertise or effort.

\subsection{Spread: Where Is the Change Distributed?}
\label{sec:axis-spread}
Spread describes how an edit is distributed across files and directories. Consider two illustrative patches, each changing 100 lines in two files. One distributes the lines 90/10 and the other 50/50. The second distributes its edits more evenly, but either patch could affect nearby or distant files. We measure edit distribution and directory coverage separately.

\paragraph{Normalized entropy}
Change entropy has been studied in relation to software faults~\cite{hassan2009predicting}. We adapt it to a single patch to measure how evenly its changed lines are distributed. For a patch $\mathcal{P}$ touching $n$ files, let $\Delta_k$ be the number of added and deleted lines in file $k$, and let $p_k=\Delta_k/\sum_{f\in\mathcal{P}}\Delta_f$. We compute
\[
H_n(\mathcal{P})=-\frac{1}{\log_2 n}\sum_{k=1}^{n}p_k\log_2 p_k.
\]
We set $H_n=0$ for a single-file patch. The value equals one when edits are evenly distributed across two or more files. The normalization allows comparison across different file counts, but it also means that two equally edited files and ten equally edited files both score one. Entropy must therefore be interpreted with the file count and other indicators.

\paragraph{Radius}
Following work on the locations of patch hunks~\cite{hunk-analysis-2025-nashid}, we measure the fraction of repository directories represented by the changed files' ancestor paths. For a file $f$ at path $d_1/\cdots/d_k/f$, define
\[
\mathrm{Anc}(f)=\{d_1,d_1/d_2,\ldots,d_1/\cdots/d_k\}.
\]
Let $D_{\mathrm{repo}}$ be the set of directory nodes in the repository. We compute
\[
R(\mathcal{P})=\frac{|\bigcup_{f\in\mathcal{P}}\mathrm{Anc}(f)|}{|D_{\mathrm{repo}}|}.
\]
Radius measures directory coverage, not pairwise distance. It can increase by reaching more directories or deeper paths, complementing entropy's measure of edit distribution.

\subsection{Novelty: How Much Code Is Added or Removed?}
\label{sec:axis-novelty}
Novelty measures the balance between additions and deletions in the diff against the base commit:
\[
\mathrm{Novelty}(\mathcal{P})=
\frac{|\mathrm{add}(\mathcal{P})|}{|\mathrm{add}(\mathcal{P})|+|\mathrm{del}(\mathcal{P})|}.
\]
A patch that only adds code scores one. A patch that only removes code scores zero, and a replacement with equal additions and deletions scores 0.5. Here, novelty refers to the form of the edit, not whether the functionality or algorithm is new. An added comment and a newly implemented function both contribute added lines. This matters when reference patches include substantial documentation.

\subsection{Centrality: What Code Does the Change Affect?}
\label{sec:axis-depth}
Centrality groups four indicators of the affected code: its incoming and outgoing dependencies, recent modification history, and callable complexity. A \emph{callable} is a function or method. These indicators describe different aspects of the code a developer may need to understand. The group name does not imply that all four are measures of graph centrality.

\paragraph{Fan-in and fan-out}
An import graph represents dependencies between modules. Fan-in counts how many modules import a module, while fan-out counts how many modules it imports. Let $\mathcal{M}(\mathcal{P})$ contain the modules with callables modified by a patch, and let $\mathcal{M}_{\mathrm{repo}}$ contain all repository modules. With $\mathrm{in}(M)$ denoting the incoming import count, we compute
\[
\mathrm{FanIn}(\mathcal{P})=
\frac{\sum_{M\in\mathcal{M}(\mathcal{P})}\mathrm{in}(M)}
{\sum_{M'\in\mathcal{M}_{\mathrm{repo}}}\mathrm{in}(M')}.
\]
Fan-out uses the same calculation with outgoing import counts. These ratios describe the share of repository import relationships associated with the affected modules. They do not establish whether any specific dependent module needs to change.

\paragraph{Churn}
Prior work uses code churn to characterize change activity~\cite{code-churn-1998-munson}. Our indicator measures how frequently recent commits touched the affected files. Let $S$ be the files modified by a patch at base commit $c_i$. We use the ancestor commits within 180 days of $c_i$, denoted by $\mathcal{C}_{180}(c_i)$, using committer dates. This window follows prior studies~\cite{churn-180-shrikanth2021early,churn-180-zimmermann2007predicting}. We compute
\[
\mathrm{Churn}(\mathcal{P})=
\frac{|\{c\in\mathcal{C}_{180}(c_i):\mathrm{files}(c)\cap S\neq\emptyset\}|}
{|\mathcal{C}_{180}(c_i)|}.
\]
A high value means that the patch affects files involved in a large fraction of recent commits. It does not measure the number of lines changed in those commits or establish that the code is unstable.

\paragraph{Mass}
We adopt the callable mass definition from~\cite{slopcodebench-2026-orlanski}, which combines size in source lines with cyclomatic complexity~\cite{mccabe-cc-1976}. Let $\mathcal{F}(\mathcal{P})$ contain the modified callables and $\mathcal{F}_{\mathrm{repo}}$ contain all repository callables. We aggregate the callable measure as
\[
\mathrm{Mass}(\mathcal{P})=
\frac{\sum_{f\in\mathcal{F}(\mathcal{P})}\mathrm{mass}(f)}
{\sum_{f'\in\mathcal{F}_{\mathrm{repo}}}\mathrm{mass}(f')}.
\]
This ratio represents the share of repository mass associated with the changed callables. Unlike a changed-line count, it accounts for the surrounding callable even when a patch edits only a small part of it.

\subsection{Comparing Benchmark Profiles}
\begin{figure}[t]
  \centering
  \includegraphics[width=\linewidth]{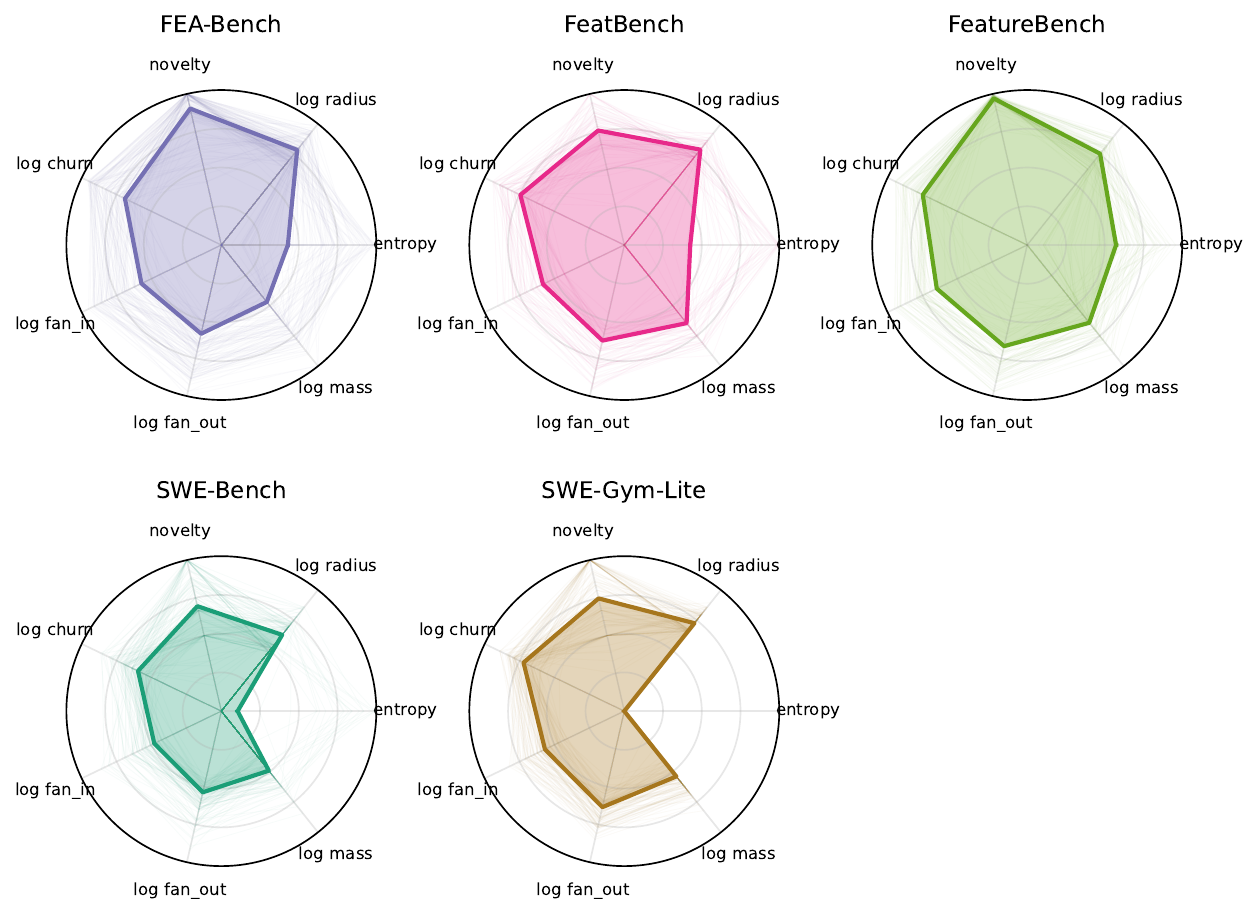}
  \caption{The seven SNC indicators for each benchmark. Faint lines represent individual tasks and bold polygons represent medians. The profiles retain differences hidden by broad task labels.}
  \label{fig:rq1-radar}
\end{figure}

For each task $i$, we compute the seven indicators from its gold patch $\mathcal{P}^{*}_i$. Figure~\ref{fig:rq1-radar} displays these profiles. For the statistical comparison, we also construct one scalar per axis. Novelty enters directly. Centrality is the unweighted mean of the $\log_{10}$ values of its four indicators. Spread uses log-transformed radius alone. We retain entropy in the detailed profile, but exclude it from this aggregate because its value is fixed at zero for single-file patches, which dominate SWE-bench and SWE-Gym Lite. We standardize each axis across the pooled tasks using z-scores and then map its values to $[0,1]$ with min--max scaling. These values describe relative positions in the corpus, not absolute difficulty thresholds.

\begin{figure*}[t]
  \centering
  \includegraphics[width=\linewidth]{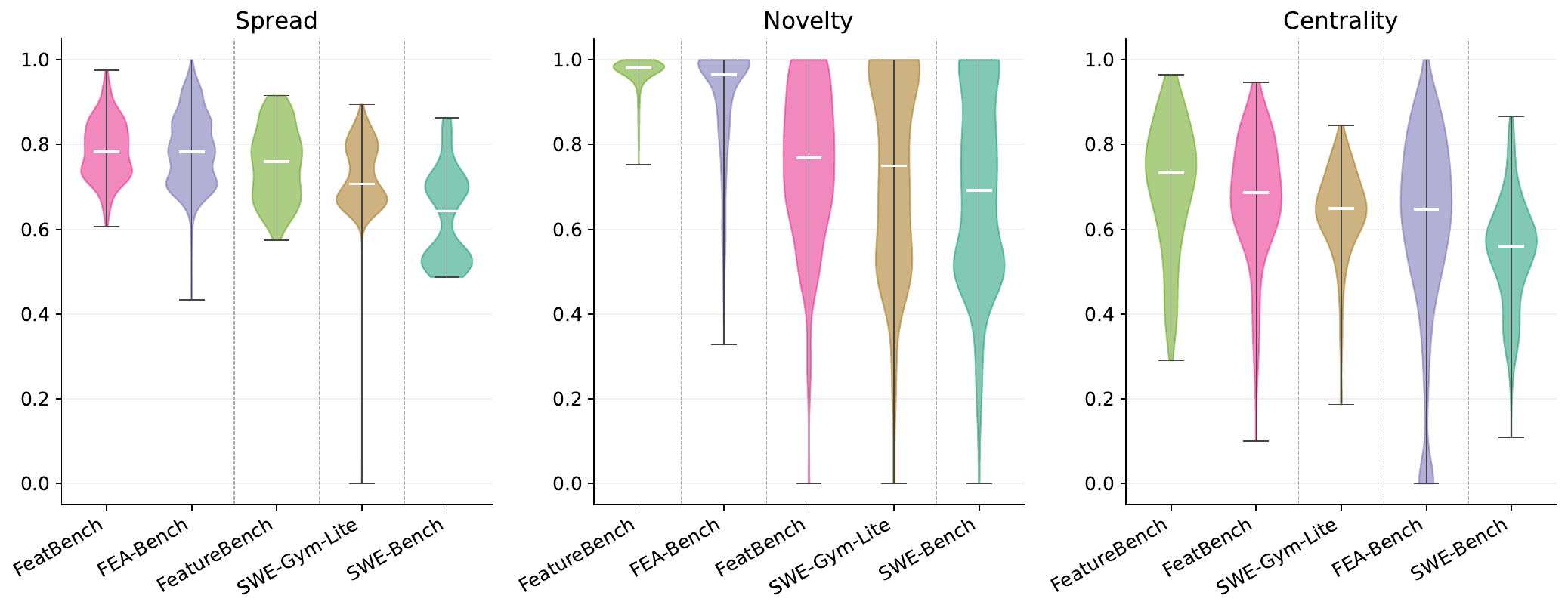}
  \caption{Benchmark distributions and Scott--Knott ESD groups on the aggregated SNC axes. Dashed separators mark group boundaries. Values are scaled relative to the pooled tasks. Spread uses radius; Centrality combines four indicators.}
  \label{fig:rq1-sk}
\end{figure*}

We compare benchmarks using the Scott--Knott Effect Size Difference (ESD) procedure~\cite{sk-1974,sk-esd-2017-tantithamthavorn}. The procedure groups benchmarks using a significance level of $\alpha=0.05$ and an effect-size threshold of $|d|\geq0.2$. Benchmarks assigned to the same group are not separated by this procedure. Figure~\ref{fig:rq1-sk} shows the groups on each aggregated axis.


\subsection{Results}
\paragraph{Observation 1: Every benchmark pair differs on at least two axes}
The procedure identifies four groups on each of the three SNC axes. The membership changes across axes, so every pair of benchmarks is separated on at least two. SWE-bench and SWE-Gym Lite differ on all three axes despite both being dominated by issue-resolution tasks. The three feature benchmarks never all occupy the same group. Thus, a shared task label is insufficient evidence that two benchmarks cover similar changes.

\paragraph{Observation 2: Construction rules help explain the differences}
FeatBench allows modifications to existing functions but excludes the addition or deletion of functions. Its Novelty distribution falls in the same group as SWE-Gym Lite and below the other two feature benchmarks. This is consistent with a collection that emphasizes changes within existing implementations.

FeatureBench's Novelty values concentrate near one. Its tasks remove code that the reference solution restores, making the reference changes predominantly additive. A high Novelty value in this setting reflects how the task was constructed. It should not be interpreted as evidence that the agent must invent an equivalent amount of new functionality.

FEA-Bench has the lowest aggregate Centrality among the three feature benchmarks. Its selection of PRs that add components provides a plausible explanation: newly introduced code can have limited history and fewer established dependencies. The aggregate result does not imply that every added component lacks incoming dependencies or that all FEA-Bench changes are simple.

\paragraph{Practical implication}
Before treating benchmarks as substitutes, compare their change distributions. Computing the same indicators on completed local changes can reveal gaps in benchmark coverage. The comparison does not by itself predict success on the local workload.

\section{RQ2: How Do an Agent's Patches and Exploration Differ Across Benchmarks?}
\label{sec:rq2}
\subsection{Motivation}
Agents may pass tests by changing different files or different amounts of code from the gold patch. We compare these patches and the files accessed during execution to describe what an evaluator receives beyond the test outcome.

We focus on the 1{,}083 tasks resolved by Claude Opus 4.6 through Claude Code. Restricting the analysis to resolved runs lets us compare patches that the benchmark accepts. It does not make the patches equivalent in maintainability, nor does it hold task difficulty constant across benchmarks: each benchmark contributes only the subset that this configuration resolves.

\subsection{Approach}
For task $i$, let $\mathcal{P}^{A}_i$ denote the final agent patch and $\mathcal{T}^{A}_i$ its trajectory. We compute three ratios for patch scope and one for exploration. We use a logarithmic scale to display reductions and increases relative to the reference on either side of zero.

\paragraph{Patch scope}
We count distinct modified files, added plus deleted lines, and distinct modified callables. For each measure $m$, we compute
\[
\rho_m(i)=\log_{10}\frac{m(\mathcal{P}^{A}_i)}{m(\mathcal{P}^{*}_i)},
\quad m\in\{\mathrm{files},\mathrm{lines},\mathrm{callables}\}.
\]
Zero means that the two patches have equal counts, which we call \emph{scope parity}. A positive value means that the agent patch is larger on that measure. For example, modifying four files when the gold patch modifies two gives a ratio of two, or approximately 0.30 on the logarithmic scale. Equal counts do not imply that the patches change the same files or implement the same design.

\paragraph{Exploration breadth}
We count the distinct files read or edited during a trajectory and compare this count with the files in the final patch:
\[
\rho_{\mathrm{explore}}(i)=\log_{10}
\frac{\mathrm{files}(\mathcal{T}^{A}_i)}{\mathrm{files}(\mathcal{P}^{A}_i)}.
\]
For example, reading or editing 20 files and retaining changes in four produces a ratio of five. This measure describes the breadth of file access relative to the final patch. It does not measure time spent reading, repeated accesses, token consumption, or the usefulness of the inspected context. Unlike the patch ratios, its denominator is the agent patch rather than the gold patch. Under complete event capture, every modified file is also a touched file, so this ratio cannot be below one before taking the logarithm.

\begin{figure*}[t]
  \centering
  \subfloat[Modified files: agent/gold.]{\includegraphics[width=.45\linewidth]{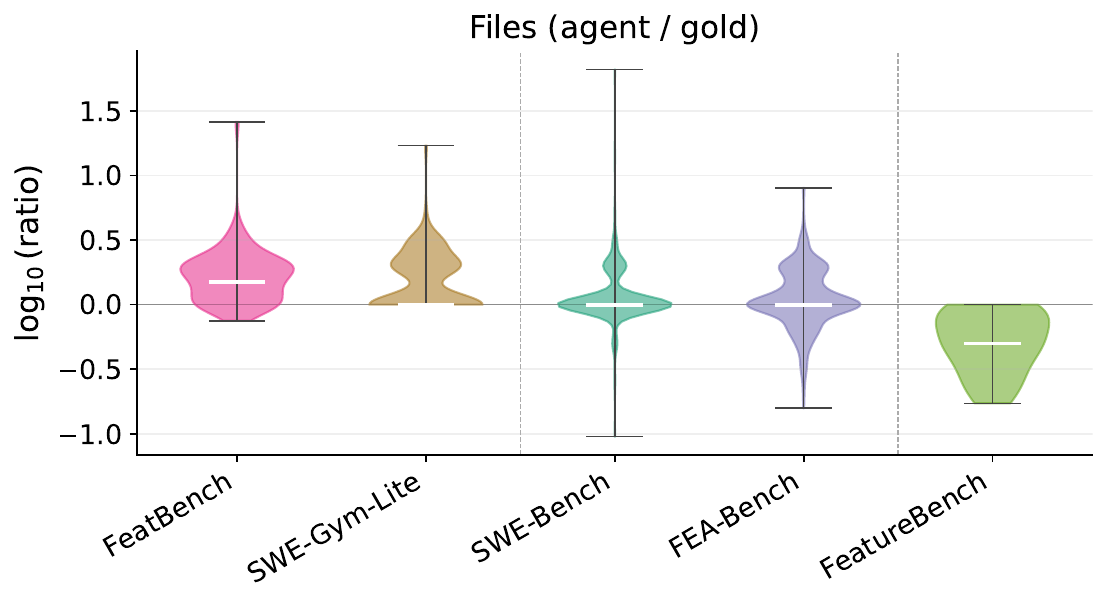}}\hfill
  \subfloat[Changed lines: agent/gold.]{\includegraphics[width=.45\linewidth]{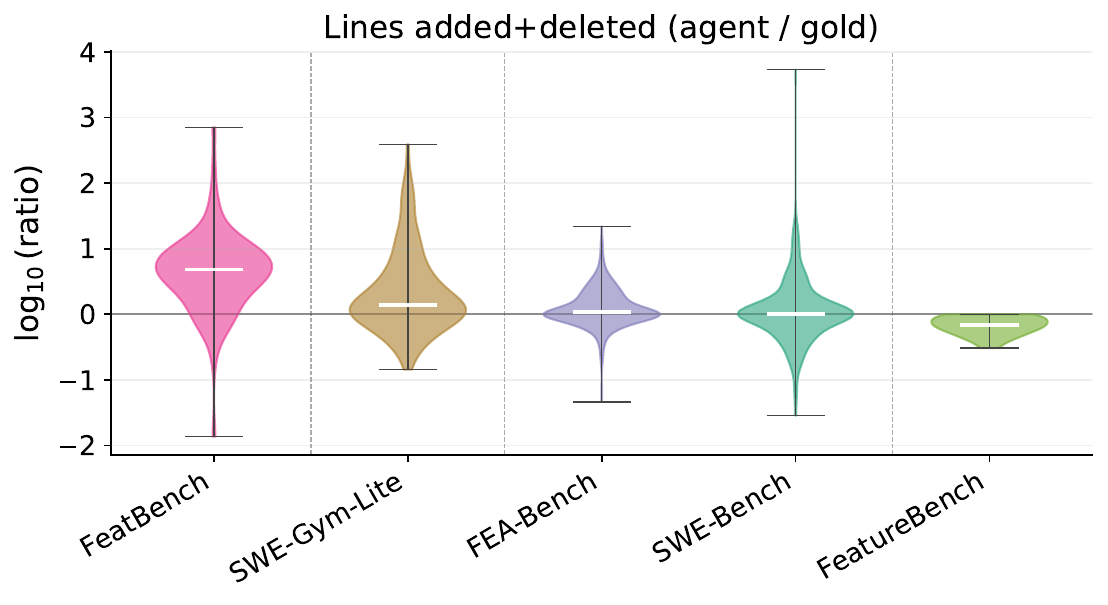}}

  \subfloat[Modified callables: agent/gold.]{\includegraphics[width=.45\linewidth]{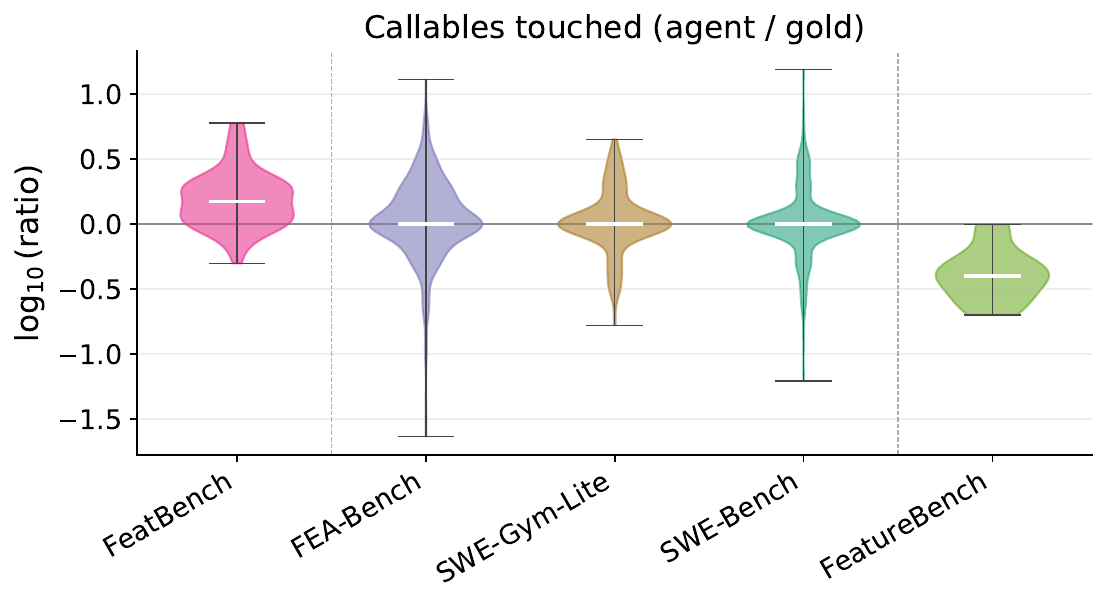}}\hfill
  \subfloat[Files accessed/final patch files.]{\includegraphics[width=.45\linewidth]{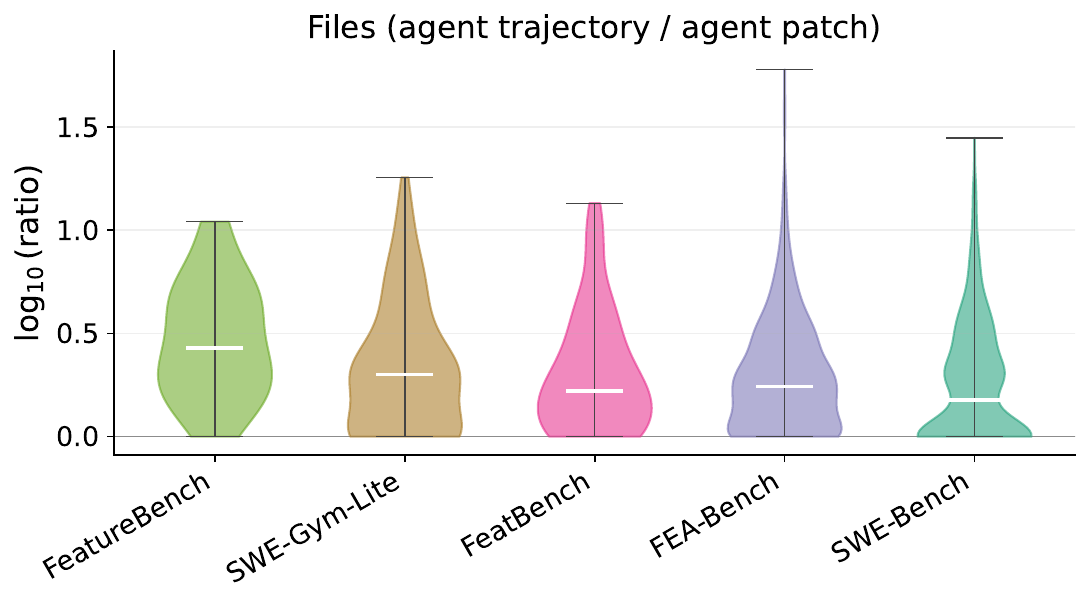}}
  \caption{Patch scope and exploration for the 1{,}083 tasks Claude Opus 4.6 resolves. Ratios are shown on a $\log_{10}$ scale; zero denotes equal counts. Dashed separators identify Scott--Knott ESD groups. All five benchmarks share one group for exploration breadth.}
  \label{fig:rq2-main}
\end{figure*}

We use the same Scott--Knott ESD procedure as in RQ1 to compare the four ratios across benchmarks. Figure~\ref{fig:rq2-main} includes all four measurements.

\subsection{Results}
\paragraph{Observation 3: Resolved patches differ systematically in scope across benchmarks}
On all three patch measures, FeatBench has the largest agent-to-gold ratios and FeatureBench has the smallest. Median log-ratios for the other three benchmarks are near zero. The reference patch therefore provides a different baseline for the same agent depending on the benchmark.

FeatBench gives the agent no implementation hints and its accepted patches exceed the gold patch in files, lines, and callables. FEA-Bench supplies signatures and docstrings and has median ratios near parity. The contrast is consistent with hints narrowing the implementation choices available to the agent. However, these benchmarks also differ in repositories, tasks, and construction rules. We did not remove or add hints to the same tasks, so this comparison does not isolate an effect of wording.

On FeatureBench, the agent passes tests with smaller patches than the reference. Comments and docstrings account for 33.8\% of the gold-patch lines in FeatureBench, and their presence provides one possible explanation for a gap in line counts, since restoring all documentation may be unnecessary for passing the tests. It does not fully explain the differences in files and callables. Smaller accepted patches could also reflect alternative implementations or incomplete test coverage.

\paragraph{Observation 4: The comparison does not separate exploration breadth across benchmarks}
All five benchmarks occupy one Scott--Knott ESD group for exploration breadth. We detect no separation under this procedure, but this does not establish identical exploration. Similar ratios can conceal different absolute file counts, and the comparison covers only one configuration's resolved runs.

\paragraph{Practical implication}
Record patch scope alongside test outcomes and inspect deviations from the reference. Document whether benchmark hints match the information available in the intended workflow. Smaller patches and narrower exploration should not be treated as universal quality goals.

\section{RQ3: Which Task and Behaviour Characteristics Are Associated with Resolution?}
\label{sec:rq3}
\subsection{Motivation}
We include resolved and unresolved runs from all six configurations to identify recurring associations between task characteristics, agent behaviour, and outcomes. These associations can guide investigation during evaluation but do not predict individual outcomes.

\subsection{Approach}
\label{sec:rq3:approach}
For each configuration, we contrast resolved and unresolved runs across the pooled benchmarks. We divide each SNC indicator into up to six quantile bins. Adjacent cut points with identical values are merged, so an indicator with many repeated values can have fewer bins. We divide the unlogged behaviour ratios into five bands: below $0.5\times$, $0.5$--$0.8\times$, $0.8$--$1.25\times$, $1.25$--$2\times$, and above $2\times$. The middle band represents approximate scope parity. For exploration, parity means that the number of accessed files is close to the number of files in the final agent patch.

\begin{figure*}[t]
  \centering
  \includegraphics[width=.85\linewidth]{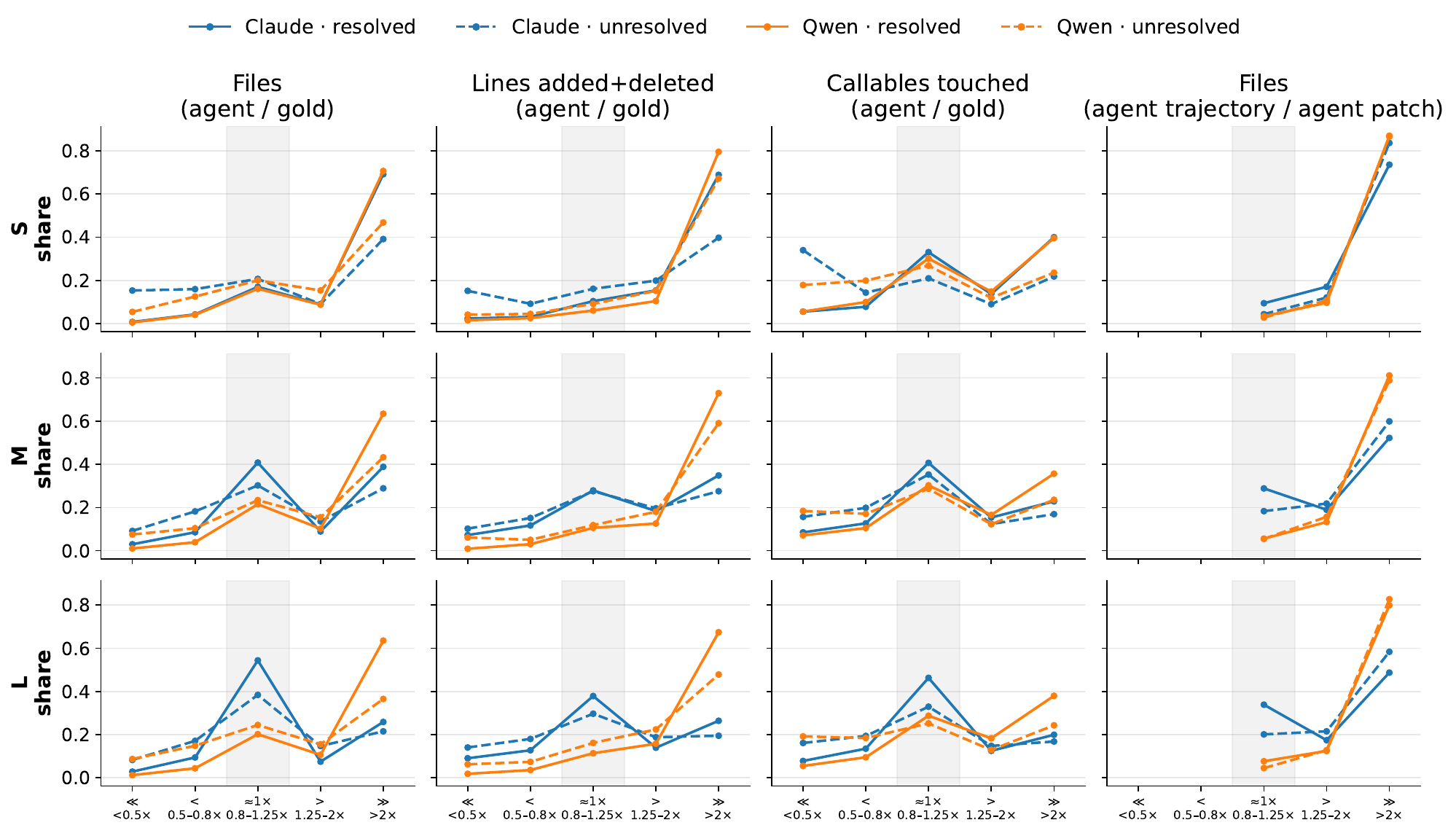}
  \caption{Behaviour distributions within resolved and unresolved runs. The shaded band represents ratios from $0.8$ to $1.25$. The first three columns compare agent and gold patches; the last compares accessed files with files in the agent's final patch. Rows show S, M, and L configurations.}
  \label{fig:rq3:behaviour}
\end{figure*}

We compare the distribution of each indicator \emph{within an outcome group}. Figure~\ref{fig:rq3:behaviour} displays the behaviour distributions. A point on the resolved curve is the share of resolved runs in that bin, not the probability of resolving a task in the bin. This distinction matters for practice. A concentration of resolved runs among less distributed changes does not establish a high success rate for those changes, because the numbers of attempted and unresolved tasks also matter.

We test the association between outcome and each binned indicator using a $\chi^2$ test of independence. We apply the Benjamini--Hochberg false discovery rate correction separately to the SNC and behaviour test families~\cite{bh-adjustment-1995}. We report Cram\'er's $V$ as an effect-size measure~\cite{cramer1946mathematical}; larger values indicate stronger associations. Table~\ref{tab:separation-stats} summarizes the effect sizes from the detailed analysis. These tests examine marginal associations, without adjusting for repository or benchmark composition.

\begin{table*}[t]
\centering
\caption{Association between outcome and binned indicators: Cram\'er's $V$. All entries have BH-adjusted $p<0.05$ except the two marked ``ns''. The SNC and behaviour blocks are separate multiple-testing families. Values reproduce the original detailed tests; larger values indicate stronger associations.}
\label{tab:separation-stats}
\begin{tabular}{@{}lrrrrrrrrrrr@{}}
\toprule
& \multicolumn{7}{c}{Gold-patch SNC indicators} & \multicolumn{4}{c}{Agent behaviour ratios} \\
\cmidrule(lr){2-8}\cmidrule(lr){9-12}
Configuration & Entropy & Radius & Novelty & Churn & Fan-in & Fan-out & Mass & Files & Lines & Calls & Explore \\
\midrule
Claude Haiku & .35 & .24 & .22 & .25 & .22 & .24 & .21 & .38 & .33 & .40 & .13 \\
Claude Sonnet & .23 & .15 & .15 & .19 & .16 & .18 & .17 & .23 & .10 & .16 & .12 \\
Claude Opus & .28 & .20 & .17 & .21 & .19 & .19 & .21 & .23 & .15 & .18 & .15 \\
Qwen 9B & .26 & .16 & .19 & .22 & .15 & .18 & .17 & .24 & .13 & .23 & .02\textsuperscript{ns} \\
Qwen 27B & .23 & .15 & .16 & .19 & .14 & .16 & .15 & .24 & .18 & .21 & .03\textsuperscript{ns} \\
Qwen 397B-A17B & .30 & .21 & .19 & .26 & .20 & .21 & .21 & .31 & .21 & .26 & .07 \\
\bottomrule
\end{tabular}
\end{table*}

\subsection{Results}
\paragraph{Observation 5: Task characteristics have similar associations across configurations}
Resolved runs tend to concentrate in lower bins of Spread and Centrality, while unresolved runs have more mass in higher bins. The broad pattern appears in both families at all three configuration levels. All 42 combinations of SNC indicator and configuration have a statistically significant association with outcome after adjustment. Their Cram\'er's $V$ values range from 0.14 to 0.35 (Table~\ref{tab:separation-stats}).

Novelty has a different pattern. Resolved runs occur at both extremes, while unresolved runs have a larger share in the middle of the range. The middle corresponds to changes that combine additions and deletions. Such changes may require adapting existing behaviour, but these distributions do not isolate that explanation from differences in the underlying tasks. In particular, a larger Novelty value should not be read as a uniformly harder task.

\paragraph{Observation 6: The patch scope associated with resolution differs by configuration}
In Figure~\ref{fig:rq3:behaviour}, Claude's resolved runs tend toward parity, particularly for callables. Qwen's resolved runs instead have approximately 0.6--0.7 of their file-count distribution above twice the gold count, with approximately 0.2 near parity. Small Claude also frequently exceeds twice the gold count for files and lines. These patterns describe the configurations as deployed, including their different agent scaffolds.

Unresolved runs generally have more mass below parity than resolved runs. A patch that is smaller than the reference is therefore a useful case to inspect in this dataset. It is not a sufficient failure criterion: RQ2 shows that accepted patches on FeatureBench are also commonly smaller than the reference. The interpretation depends on both the benchmark and the configuration.

\paragraph{Observation 7: Larger Claude configurations more often match the reference scope}
Among resolved Claude runs, the share in the file parity band increases from 0.17 for Haiku to 0.41 for Sonnet and 0.54 for Opus. For exploration, the corresponding share increases from 0.09 for Haiku to 0.34 for Opus. The resolved Qwen distributions show no comparable movement toward parity; approximately 0.8 of resolved Qwen runs remain above twice the final patch's file count in exploration at each level.

The behaviour associations are significant in 22 of 24 comparisons. Qwen 9B and 27B exploration are the exceptions, with $V=0.02$ and $0.03$. Even the significant association for Qwen 397B-A17B exploration is small ($V=0.07$). This illustrates why statistical significance alone is insufficient for choosing a diagnostic. Also, each configuration resolves a different subset of tasks. The changing distribution among resolved runs cannot establish that increasing model size makes a given task's patch smaller.

\paragraph{Practical implication}
Task profiles offer a common way to organize evaluation results across configurations. Patch scope needs a more local interpretation. Before adding automated scope checks, teams should establish the distribution for their chosen agent and workload and inspect the cases that differ. Our results motivate this assessment but establish no thresholds for rejection, escalation, or model selection.

\section{Lessons for Practitioners}
\label{sec:discussion}
Our findings inform task selection, patch assessment, and investigation of agent behaviour. We derive four lessons and a local evaluation workflow. As future work, we plan to evaluate this workflow as an industrial intervention.

\subsection{Select Evaluation Tasks by Change Characteristics}
\textbf{Lesson 1: A second benchmark with the same label can add different coverage.} Every benchmark pair differs on at least two SNC axes. Teams choosing between collections labelled feature implementation should therefore inspect their actual changes. FeatBench emphasizes modifications within existing functions, while FeatureBench emphasizes restoration of removed code. A team maintaining existing behaviour may find the former relevant, while a team evaluating implementation from specified interfaces may also need tasks with explicit signatures. The appropriate mix depends on the intended work.

The same lesson applies to training and evaluation data. A collection of many tasks can still emphasize a narrow range of changes. Teams generating tasks, for example with approaches such as SWE-smith~\cite{swesmith-2025-yang}, could use the profile to examine which kinds of changes their generator produces. Benchmark authors could publish per-task indicators and document how selection rules constrain them. These additions would complement dataset documentation~\cite{gebru2021datasheets} by making the distribution of code changes available for inspection.

\textbf{Lesson 2: The information supplied with a task is part of the evaluation setting.} The same change can be requested with or without file locations, interface signatures, or implementation guidance. RQ2 shows different patch-scope distributions for collections with different forms of guidance. Because the study does not isolate guidance experimentally, the practical response is to record it and compare it with the intended workflow. If engineers normally provide an API design before delegating implementation, interface hints may be representative. If they provide only a user-facing requirement, an evaluation that supplies detailed signatures leaves a localization and design requirement untested.

\subsection{Inspect Accepted Patches Before Defining Scope Policies}
\textbf{Lesson 3: A reference patch is a comparison point, not a prescribed patch size.} Opus resolves tasks with larger patches than the reference on FeatBench and smaller patches on FeatureBench. A scope deviation therefore needs an explanation. During evaluation, reviewers can check whether extra files implement necessary integration, whether omitted code is documentation, and whether tests permit an incomplete implementation. This inspection connects a numerical difference to an engineering consequence.

\textbf{Lesson 4: Establish a baseline for each deployed configuration.} The Claude and Qwen configurations have different patch distributions among resolved runs, and small Claude differs from larger Claude configurations. These observations argue for measuring behaviour after choosing the model and scaffold together. They do not show that imposing a smaller-diff instruction will help one family or harm another. A team considering such an instruction should compare it on the same tasks and record both test outcomes and the reasons reviewers accept or reject the resulting changes.

Patch size can help select cases for review, but does not measure review effort. A longer explicit implementation may be easier to understand than a shorter complex one. Review time, requested revisions, and subsequent defects would help establish the operational consequences.

\subsection{A Workflow for Applying the Analysis Locally}
A team can apply the analysis to completed changes in four steps.

\paragraph{Define the work the evaluation should represent}
Select historical changes from the repositories and task types intended for agent use. Retain the request, base commit, accepted patch, and available tests. Record the information developers had when starting the task. Keep cases with missing requests or tests visible as coverage gaps; excluding them silently could make the evaluation easier than the intended workload. An accepted historical patch is a reference implementation, with the same limitations as a benchmark gold patch.

\paragraph{Compare the local task profile with candidate benchmarks}
Compute the seven indicators and inspect their distributions alongside file and line counts. Look for local changes that are poorly represented, such as edits spread across several directories or changes involving modules with many dependents. Retain a common normalization when comparing collections, since independently scaling each collection can hide differences. Use the profile to choose complementary public tasks or identify local tasks worth evaluating. Similar profiles provide evidence about structural coverage, but do not establish equivalence of requirements or domain knowledge.

\paragraph{Evaluate the complete configuration and retain its evidence}
Record the model, scaffold, instructions, tools, limits, final patch, and outcome for every attempt. Preserve failures to start or finish separately from test failures so that the operational reliability of the evaluation can be examined. Report resolution rates with attempted and resolved counts for meaningful task groups. This adds the success-rate information that cannot be read directly from our outcome-conditioned plots. Retaining trajectories also makes it possible to inspect why an unusual scope ratio occurred.

\paragraph{Inspect deviations before changing the workflow}
Review examples with unusually small or large patches, including both resolved and unresolved runs. Check whether the proposed change meets the request, whether it modifies unrelated code, and what the tests cover. Use the inspected cases to formulate a specific intervention, such as clearer interface guidance or additional validation. Compare the intervention with the existing configuration on the same evaluation tasks before making it a default. This is especially important for runtime escalation rules, because the gold patch used in our analysis is normally unavailable for a new request.

\subsection{Costs, Benefits, and Evidence Still Needed}
The profile requires patches, repository structure, and commit history; exploration analysis also requires action logs. Extracting imports, matching callables, and handling incomplete histories require implementation work. We have not measured integration effort in company tooling.

The demonstrated benefit is more specific information about evaluation coverage and accepted patches. To assess industrial impact, record whether the analysis changes benchmark selection, which decisions differ from those based on aggregate scores, and whether the resulting task mix exposes previously missed failures. Productivity and review costs require separate measurement.

\section{Threats to Validity}
\label{sec:threats}
\subsection{Construct Validity}
The SNC profile measures properties of a reference change, not the full engineering work required by a request. It omits requirements ambiguity, domain knowledge, runtime interactions, and organizational constraints. Directory coverage depends on repository layout; import relationships miss dependencies that do not appear as imports. Churn describes recent activity, which need not indicate difficulty. Novelty counts added and removed lines, including changes that may not affect functionality. The aggregate axes further compress distinct indicators. We retain the individual indicators to make these limitations inspectable.

A gold patch is one accepted implementation. Counts that match it do not establish equivalent behaviour, and counts that differ do not establish a defect. Benchmark resolution is determined by the available tests. Tests can miss requirements, and the agent may produce a patch that passes without being suitable for maintenance. Our exploration measure depends on the completeness of recorded file accesses and says nothing about the usefulness of the information obtained.

\subsection{Internal and Statistical Validity}
Claude and Qwen use different scaffolds. Cross-family contrasts therefore concern deployed configurations, rather than isolated model effects. Within-family comparisons also involve different sets of resolved tasks, so a shift among resolved runs may reflect which tasks a configuration can solve. RQ2 restricts attention to one configuration's successful subset and cannot explain behaviour on all attempted tasks.

The benchmark comparisons are observational. Construction rules, repositories, hints, and reference-patch conventions vary together. We offer explanations consistent with these differences, but do not establish that hints or any single construction rule cause a behaviour change. Repeated runs would be needed to quantify stochastic variation~\cite{randomness-agentic-2026-bjarnason}.

Tasks from the same repository may be related. In addition, FEA-Bench supplies more than half of the pooled tasks, and the same tasks recur across configurations. Our marginal tests do not adjust for these dependencies or for benchmark composition. Multiple-testing correction limits false discoveries within the stated test families but does not remove these threats. We report effect sizes and avoid interpreting significant associations as causal effects or calibrated predictions. In particular, the RQ3 distributions cannot establish the reliability of delegating tasks from a given SNC region.

\subsection{External Validity}
The five collections contain Python tasks for issue resolution and feature implementation. Other languages, proprietary systems, refactoring~\cite{gautam2025refactorbench}, and changes that span several repositories may have different profiles. Public benchmark tasks also differ from work that depends on internal documentation or discussion with stakeholders. The observed behaviour applies to the studied configurations and versions; it need not persist across model or scaffold updates. Local replication and practitioner assessment are needed before using the findings to guide task assignment, review effort, or production deployment.

\section{Conclusion}
\label{sec:conclusion}
We examined a practical limitation of coding-agent evaluation: broad task labels and aggregate scores give teams limited information about whether a benchmark represents their work. Across 2{,}487 tasks and 14{,}922 runs, the SNC profile reveals differences between every benchmark pair, while patch and exploration measurements show how accepted solutions vary across benchmarks and configurations. Our observations suggest that teams should compare evaluation coverage with completed changes, document the guidance supplied to agents, and inspect patch scope alongside test outcomes. The proposed workflow supports these assessments using repository artifacts. Practitioner evidence is needed to assess improvements in evaluation decisions and development outcomes.

\section*{Data Availability}

Code for computing the SNC profile and behavioural footprints and reproducing our analyses is available at \url{https://github.com/radinshayanfar/task_snc}.

\section*{Acknowledgment}
The authors used generative AI tools to assist with implementation and manuscript formatting and editing. The authors reviewed and take responsibility for the content.
\bibliographystyle{IEEEtran}
\bibliography{aaai2027}
\end{document}